\documentclass[fleqn,12pt,twoside]{article}
\usepackage{espcrc1}
\usepackage{amsmath}

\usepackage{graphicx}
\usepackage[figuresright]{rotating}

\title{Strong and Weak Interactions in $ B\to\pi^+\pi^-K$ Decays}

\author{B. Loiseau\address[LPNHE]{LPNHE (IN2P3--CNRS--Universit\'es Paris 6 et 7), Groupe Th\'eorie,\\ 
Universit\'e Pierre et Marie Curie, 4 place Jussieu, 75252 Paris, France}
\thanks{B. E.  and B.L. acknowledge useful discussions with J. P. Dedonder, O. Leitner and Craig Roberts.} , 
B. El-Bennich\addressmark[LPNHE], 
A. Furman\address{ul. Bronowicka 85/26, 30-091 Krak\'ow, Poland},
R. Kami\' nski\address[POL]{Division of Theoretical Physics, The Henryk Niewodnicza\'nski Institute of Nuclear Physics, Polish Academy of Sciences, 31-342 Krak\'ow, Poland},
L.~Le\'sniak\addressmark[POL]}

\begin{document}


\maketitle

\vspace{-8cm} 
\textit{Contribution to the 18th International IUPAP Conference on Few-Body Problems in Physics (FB18), Santos, Brasil, August 21-26, 2006}

\vspace{0.2cm} 
LPNHE 2006-13

\vspace{6.cm} 

\begin{abstract}
To describe the weak three-body decays $B\to\pi^+\pi^-K$, we recently derived amplitudes based on two-body QCD factorization followed by $\pi^+\pi^-$ final state interactions in isoscalar $S$- and isovector $P$-waves.
We study here the sensitivity of the results to the values of the $B$ to $f_0(980)$ transition form factor and to the effective decay constant of the $f_0(980)$. 

\end{abstract}

\section{INTRODUCTION}

It is important to understand charmless three-body $B$ decays to probe the standard model.
These decays are sensitive to $CP$ violation and supply information on strong interactions.
To interpret in a reliable way weak decay observables it is important to take into account final state interactions between produced meson pairs.
In the  weak decays $B\to\pi^+\pi^-K$~\cite{Abe05,Aub06}
one sees maxima around the $\pi\pi$ effective mass distributions in the $\rho(770)^0$ and $f_0(980)$ resonance regions.

For $\pi\pi$ effective mass $m_{\pi\pi}$ up to 1.2 GeV, the contribution of the isospin-zero $S$-wave 
$(\pi^+\pi^-)_S$ final state interactions was described in Ref.~\cite{Fur05} and that  of the isospin-one $P$ wave $(\pi^+\pi^-)_P$ was included in Ref.~\cite{ElB06}.  The amplitudes, based on the QCD factorization approach without the inclusion of hard-spectator and annihilation terms, underestimate the $B$ to $\rho(770)^0 K$ and $f_0(980) K$ branching fractions. Therefore,
phenomenological amplitudes arising from enhanced $c\bar c$ loop diagrams (charming penguin terms~\cite{Ciu97}) were added. 
Our presentation at the conference was based on the  work described in Ref.~\cite{ElB06}.
Here we show the sensitivity of the model~\cite{ElB06} to  two inputs of the $S$-wave amplitude: the $B$ to $f_0(980)$ transition form factor and  the effective decay constant of the $f_0(980)$.

\section{WEAK DECAY AMPLITUDES FOR $\boldsymbol{ B\to\pi^+\pi^-K}$}

The amplitudes for the weak decays $B\to(\pi^+\pi^-)_{S(P)}K$ are derived~\cite{Fur05,ElB06} in the QCD factorization framework~\cite{Buc96,bene03}. As a first approximation, corrections arising from annihilation topologies and hard gluon scattering with the spectator quark are not included.
These also contain several phenomenological parameters (see for instance~\cite{bene03}).

For the $B\to(\pi^+\pi^-)_SK$ decay amplitudes, we
consider the three-body $\pi^+\pi^-K$ final state as arising from a quasi two-body one with the produced $\pi^+\pi^-$ pair being in an isospin-0 $S$-wave state $R_S$ of mass $m_{\pi\pi}$.
For $m_{\pi\pi}=m_{f_0}$  (mass of the $f_0(980)$) this $R_S$ state is the $f_0(980)$ resonance.
Our amplitudes have a weak two-body decay part based on operator product expansion, heavy quark limit and QCD factorization~\cite{Buc96} followed by the strong decay of $R_S$ into $(\pi^+\pi^-)_S$ with inclusion of rescattering.
One has~\cite{Fur05,ElB06},
\begin{equation}
\label{eq:1}
 \langle(\pi^+\pi^-)_SK^-\vert H\vert B^-\rangle=
 M_K^S(m_{\pi\pi})+M_{R_S}^S(m_{\pi\pi})\ ,
\end{equation}
\begin{equation}
\begin{array}{l}
\label{eq:2}
M_K^S(m_{\pi\pi})=
\displaystyle\frac{G_F}{\sqrt{2}}f_K(M_B^2-m_{\pi\pi}^2)F_0^{B\to R_S}(M_K^2)
\bigg\{
\lambda_u\left[a_1+a_4^{u}-a_4^c+\left(a_6^c-a_6^{u}\right)r-S_u\right]
\\
\hfill
+\lambda_t\left(a_6^cr-a_4^c-S_t\right)
\bigg\} 
\sqrt{\displaystyle\frac{2}{3}}\ \Gamma_{R_S\pi\pi}^n(m_{\pi\pi})\ ,
\end{array}
\end{equation}
\begin{equation}
\begin{array}{l}
\label{eq:3}
M_{R_S}^S(m_{\pi\pi})=
\displaystyle\frac{G_F}{\sqrt{2}}(M_B^2-M_{K}^2)
\Bigg\{
\langle R_S\vert\bar ss\vert 0\rangle
\frac{2F_0^{B \to K}(m_{\pi\pi}^2)}{m_b-m_s}
\left[
\lambda_u\left(a_6^c-a_6^{u}\right)+\lambda_ta_6^c
\right]\\
\hfill - N_K(\lambda_uS_u+\lambda_tS_t)
\Bigg\}
\sqrt{\displaystyle\frac{2}{3}}\ \Gamma_{R_S\pi\pi}^s(m_{\pi\pi})\ .
\end{array}
\end{equation}
\begin{figure}
\includegraphics[width=\textwidth,bb= 50 620 595 740]{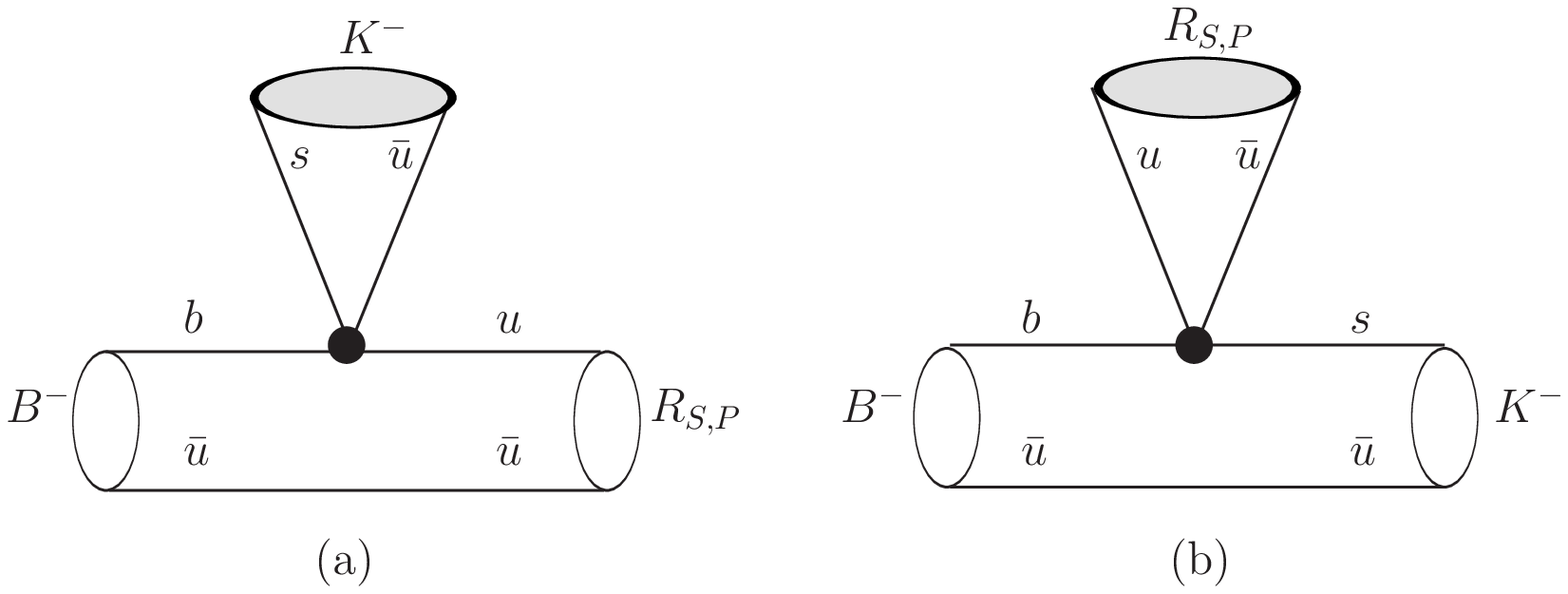}
\caption{Quark-line diagrams for two-body $B^-$ decay into $K^-$ and a $\pi\pi$ state $R_S(m_{\pi\pi})$ or $R_P(m_{\pi\pi})$ in an isoscalar $S$- or isovector $P$-wave, respectively. The filled circle represents the weak and electroweak decays via tree or penguin type diagrams.}
\end{figure}

In Eqs.~(\ref{eq:2}) and (\ref{eq:3}), $G_F$ is the Fermi  constant,  $f_K$ the kaon decay constant and $M_B,\  M_K,\  m_b$ and $m_s$ the $B$-meson, kaon, $b$- and $s$-quark masses, respectively.
 The functions $F_0^{B\to R_S}(M_K^2)$ and
$F_0^{B\to K}(m_{\pi\pi}^2)$ are the $B^-$ to $R_S$ or $K^-$ transition form factors.
 The $\lambda_{u,t}$ are products of the CKM matrix elements, 
$\lambda_{u}= V_{ub}{V}^*_{us}$ and  $\lambda_{t}= V_{tb}{V}^*_{ts}$. The $a_i$ are the scale dependent effective coefficients built from the Wilson coefficients and including next-to-leading-order QCD corrections~\cite{Buc96,bene03}. The chiral factor $r=2M_K^2/[(m_b+m_u)(m_s+m_u)]$, $m_u$ being the $u$-quark mass. The phenomenological charming penguin parameters $S_u$ and $S_t$ are added to the QCD factorization terms. In Eq. (3) the weight factor  $N_K$ is chosen to be $N_K=f_KF_0^{B\to R_S}(M_K^2)$ as in Eq. (2).

The amplitude $M_K^S(m_{\pi\pi})$  matches the topology of Figure 1 (a)  with the production of a $K^-$ meson from the vacuum plus a $B^-$ transition into an $R_S(m_{\pi\pi})$ state. The amplitude $M_{R_S}^S(m_{\pi\pi})$ corresponds to the topology of  Figure 1(b) with the emission of  an 
$R_S(m_{\pi\pi})$ state from the vacuum plus a $B^-$ transition into a $K^-$ meson.

 In Eqs.~(\ref{eq:2}) and (\ref{eq:3}), the non-strange $\Gamma_{R_S\pi\pi}^n(m_{\pi\pi})$ and strange 
$\Gamma_{R_S\pi\pi}^s(m_{\pi\pi})$ vertex functions describe the strong decay of the state $R_S(m_{\pi\pi})$ into two pions and include $\pi^+\pi^-$ rescattering~\cite{Fur05}. One can write
\begin{equation}
\label{eq:4}
\langle (\pi\pi)_S\vert \bar ss\vert 0\rangle=\Gamma_{R_S\pi\pi}^s(m_{\pi\pi})\langle R_S\vert \bar ss\vert 0\rangle
=\sqrt{2}B_0{\Gamma_1^{s}}^*(m_{\pi\pi})
\end{equation}
where $\Gamma_1^s(m_{\pi\pi})=\langle 0\vert \bar ss\vert (\pi \pi)_S\rangle/(\sqrt{2}B_0)$ is the strange scalar form factor and the normalization constant
$B_0=-\langle 0\vert \bar qq\vert 0\rangle/f_\pi^2$, $f_\pi$ being the pion decay constant.
Replacing $\bar ss$  by $\bar nn=(\bar uu+\bar dd)/\sqrt{2}$ in Eq.~(\ref{eq:4}) gives 
the non-strange vertex function $\Gamma_{R_S\pi\pi}^n(m_{\pi\pi})$ in terms of the non-strange scalar form factor 
$\Gamma_1^n(m_{\pi\pi})$.
Defining a scalar decay constant $f_{R_S}^s$ by
\begin{equation}
\label{eq:5}
\langle R_S\vert \bar ss\vert 0\rangle=m_{R_S}f_{R_S}^s,
\end{equation}
one then obtains from Eq.~(\ref{eq:4}), 
\begin{equation}
\label{eq:6}
\Gamma_{R_S\pi\pi}^s(m_{\pi\pi})=\chi{\Gamma_1^s}^*(m_{\pi\pi})\ \mbox{with}\
\chi=\sqrt{2}B_0/\left(m_{R_S}f_{R_S}^s\right).
\end{equation}
If we assume $\Gamma_{R_S\pi\pi}^n(m_{\pi\pi})=\chi{\Gamma_1^{n}}^*(m_{\pi\pi})$ and identify $R_S(m_{f_0})$ with $ f_0(980)$,
we normalize $\chi$ by~\cite{Fur05}
\begin{equation}
\label{eq:8}
\chi=g_{f_0\pi\pi}/[m_{f_0}\Gamma_{\mbox{tot}}(f_0)\vert\Gamma_1^n(m_{f_0})],
\end{equation}
where  $\Gamma_{\mbox{\rm{tot}}}(f_0)$ is the total $f_0(980)$ width.

 Note that
 replacing 
$ \sqrt{2/3}\ \Gamma_{R_S\pi\pi}^{n,s}(m_{\pi\pi})$ by 1 in Eqs.~(\ref{eq:2}) and 
(\ref{eq:3}) leads to a two-body $B^-\to R_SK^-$ decay amplitude.

In the $B\to(\pi^+\pi^-)_PK$ decay amplitudes,
the produced $\pi^+\pi^-$ pair is in an isovector $P$-wave $(\pi^+\pi^-)_P$ state 
$R_P(m_{\pi\pi})$ identified as the $\rho(770)^0$ resonance.
The explicit expression of the $B^-\to(\pi^+\pi^-)_PK^-$  amplitude is given in Ref.~\cite{ElB06}. As the amplitudes underestimate the $B$ to $\rho(770)^0 K$ branching fraction, 
we also introduce  a phenomenological charming penguin term depending on two complex parameters proportional to $\lambda_u$ and $\lambda_t$.

The complete $B^-\to(\pi^+\pi^-)_{S+P}K^-$ amplitude is obtained by adding the $S$-wave amplitude of Eq.~(\ref{eq:1}) to that of the  $P$-wave~\cite{ElB06}.
Replacing $\lambda_u,\ \lambda_t$ by $\lambda_u^*,\ \lambda_t^*$ gives the
$B^+\to(\pi^+\pi^-)_{S}K^+$ amplitude.
The expressions for the  neutral $B$-decay and $B^+\to(\pi^+\pi^-)_{P}K^+$  amplitudes can be found in~\cite{Fur05} and~\cite{ElB06}. In the results shown below we use the same input parameters as in Ref.~\cite{ElB06} unless otherwise stated.

\section{FIT, RESULTS, DISCUSSIONS AND CONCLUSIONS}

In Ref.~\cite{ElB06}, we use 
$F_0^{B\to(\pi\pi)_S}(M_K^2)=0.46$ although recent calculations~\cite{Cheng:2005nb,Leitner:2006hu,El-Bennich:2006yy} give a value close to 0.25. Furthermore,
the $\Gamma_1^{n,s}(m_{\pi\pi})$ depend on some poorly determined low energy constants of  chiral perturbation theory.
Using their latest determinations~\cite {laehde06}, the moduli of the $\Gamma_1^{n,s}(m_{\pi\pi})$ are larger by a factor of 1.25 than those of~\cite{ElB06} in the $f_0(980)$ range.
However, the constant value of $\chi\vert\Gamma_1^n(m_{f_0})\vert$ (see Eq.~(\ref{eq:8})) limits the sensitivity to this variation.
With these new inputs for the $S$-wave, the fit to the Belle and BaBar collaboration data (see references in~\cite{ElB06}) is similar to that of Ref.~\cite{ElB06}. The $\chi^2/d.o.f.=346/(222-8)=1.62$ in model~\cite{ElB06} and 1.65 here. In this new fit, the resulting charming penguin parameters are modified and in particular,
to compensate the decrease of $F_0^{B\to (\pi\pi)_S}(M_K^2)$ from 0.46 to 0.25, the modulus of $S_u$ increases.

\begin{figure}[h]
\includegraphics*[width=\textwidth,bb= 36 500 567 730]{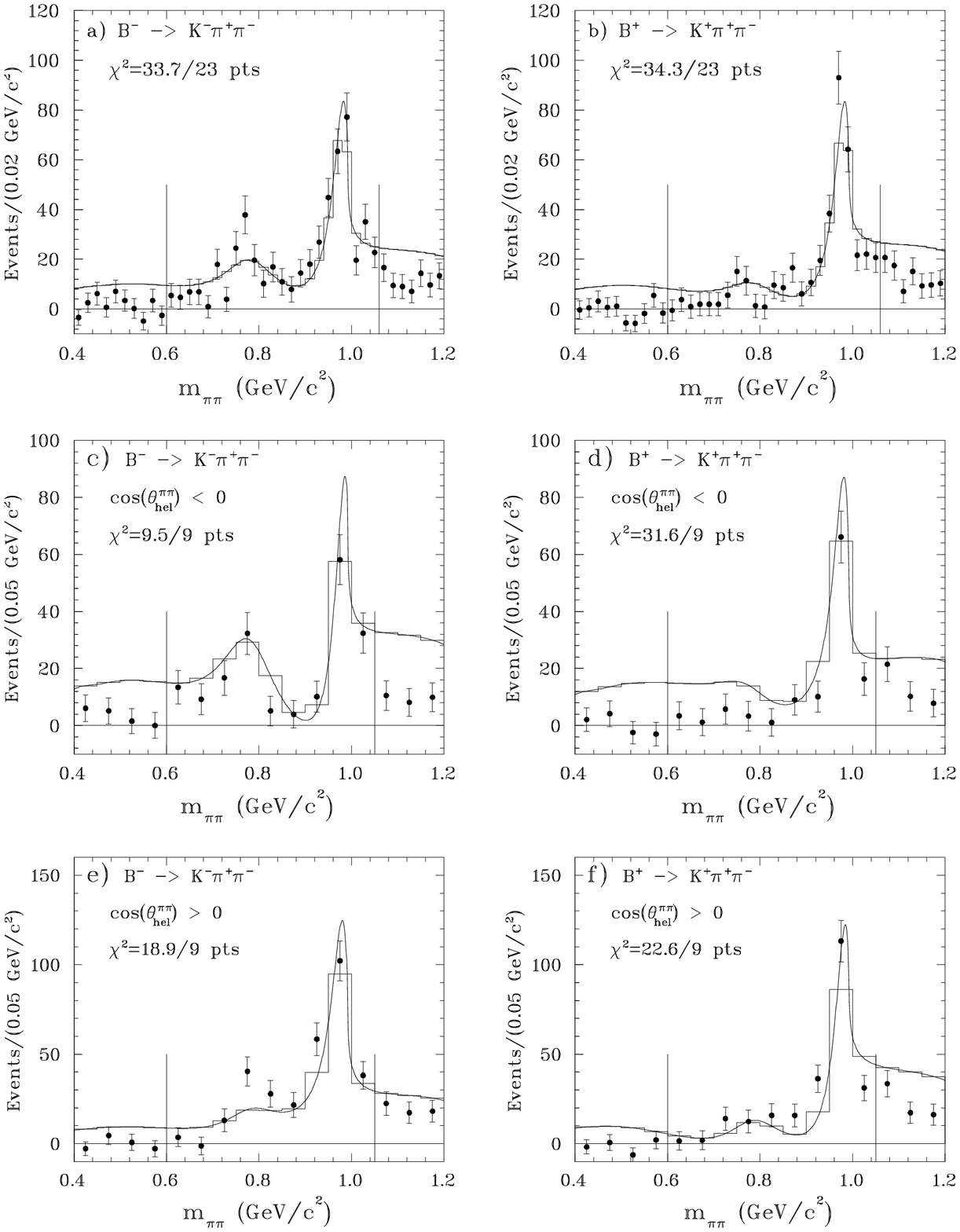}
\caption{The $m_{\pi\pi}$  distributions in $B^\pm\to\pi^+\pi^-K^\pm$ decays (data from Ref. \protect\cite{Abe05}). The solid line results from the fit with $F_0^{B\to(\pi\pi)_S}(M_K^2)=0.25$ and with the $\Gamma_{1}^{n,s}(m_{\pi\pi})$ of larger moduli in the $f_0(980)$ range (see text). The vertical lines delimit the region of the fit.}
\end{figure}

In Fig. 2 the $m_{\pi\pi}$ branching fraction distributions are compared to the Belle data~\cite{Abe05}. One sees an asymmetry in the number of events between the 
$B^-\to\pi^+\pi^-K^-$ and $B^+\to\pi^+\pi^-K^+$ decays for the $\rho(770)^0$ and $f_0(980)$ regions.
This results in a large direct $CP$ asymmetry for $B^\pm\to\rho(770)^0K^\pm$ decays of $0.32\pm 0.03$ to be compared with $0.30\pm0.14$ and $0.32\pm0.16$ from Belle and BaBar collaborations, Refs.~[1] and [4] in~\cite{ElB06}, respectively.

It was found in Ref.~\cite{Cheng:2005nb} that the experimental average branching fractions of two-body $B^\pm\to f_0(980)K^\pm$ decays, $\mathcal{B}_{f_0}$, could be reproduced without charming penguin terms. However, 
we have important differences with Ref.~\cite{Cheng:2005nb} in the $S$-wave inputs.

First, concerning the $M_{R_S}^S(m_{\pi\pi})$ amplitude, we have shown in section 2 that, with $R_S\equiv f_0$, we use an effective scalar decay constant $f_{f_0}^s=\sqrt{2}B_0/(m_{f_0}\chi)=94$ MeV with the input parameters of Ref.~\cite{ElB06} or 117.5 MeV with the new $\Gamma_1^{n,s}$ just considered above.
These values are to be compared with those of 370 MeV determined from QCD sum rules in Ref.~\cite{Cheng:2005nb} or of 245 MeV obtained  in Ref.~\cite{Bhagwat06} applying the Dyson-Schwinger equations, which respect dynamical chiral symmetry breaking in modeling scalar mesons.

With  $F_0^{B\to(\pi\pi)_S}(M_K^2)=0.46$, if we set $S_u=S_t=0$ and $\Gamma_1^{s*}(m_{\pi\pi})=0$ (equivalent to an effective $f_{f_0}^s=0$),
$\mathcal{B}_{f_0}=2.19\ (0.67)\times 10^{-6}$ when integrating the  $m_{\pi\pi}$ average distribution from 0.900 to 1.060 GeV. Here and below, the cited number in parenthesis corresponds to the fit with $F_0^{B\to(\pi\pi)_S}(M_K^2)=0.25$ and with the $\Gamma_{1}^{n,s}(m_{\pi\pi})$ of larger moduli in the $f_0(980)$ range. With the addition of $M_{R_S}^S(m_{\pi\pi})$ of~\cite{ElB06} (effective $f_{f_0}^s=94\ (117.5)$ MeV),
$\mathcal{B}_{f_0}=2.66\ (1.09)\times 10^{-6}$.
If we multiply ${\Gamma_1^{s}}^*(m_{\pi\pi})$ by 4 (3.15) (effective $f_{f_0}^s=376\ (370)$ MeV), 
$\mathcal{B}_{f_0}=4.50\ (2.49)\times 10^{-6}$. Remind that the corresponding Belle value is $6.06\pm1.08\times 10^{-6}$.
In our case, if we add the contributions of $S_u$ and $S_t$ of Ref.~\cite{ElB06}, then $\mathcal{B}_{f_0}$ increases from 
$2.66\ (1.09)\times 10^{-6}$ to $6.93\ (6.59)\times 10^{-6}$.

Secondly, the fit to experimental data of Ref.~\cite{Cheng:2005nb} includes hard spectator scattering terms with the parameters $X_A$ (plus annihilation terms with the parameters $X_H$).
As stated in Ref.~\cite{cheng05}, the  $a_1(f_0K)$ receives a large contribution from hard spectator interaction which will enhance the
$M_K^S(m_{f_0})$ of Ref.~\cite{Cheng:2005nb}.

In summary, uncertainties in $F_0^{B\to(\pi\pi)_S}(M_K^2)$ and in the $S$-wave scalar form form factors lead to variations of charming penguin parameters, in particular of $S_u$. The scalar form factors that we use give low values for the effective $f^s_{f_0}$ decay constant equal to 94 or 117.5~MeV, to be compared to 370 MeV~\cite{Cheng:2005nb} or 245 MeV~\cite{Bhagwat06}. Despite these uncertainties, our conclusions~\cite{ElB06} are unchanged. Our theoretical model is based on quasi two-body QCD factorization followed by $S$- and $P$-wave final state interactions between the produced $\pi \pi$ pairs. These interactions are constrained by  other experiments, unitarity and chiral symmetry. Our model gives a good fit of the three-body  $B\to\pi^+\pi^-K$ decay data. In particular it describes well the interference between the $f_0(980)$ and $\rho(770)^0$ resonances.


\end{document}